\documentclass[prd,twocolumn,showpacs,amsmath,amssymb]{revtex4}

\usepackage{times}
\usepackage{graphicx}

\newcommand{\beq}{\begin{equation}}
\newcommand{\eeq}{\end{equation}}

\newcommand{\ba}{\begin{eqnarray}}
\newcommand{\ea}{\end{eqnarray}}

\def\t{\theta}

\def\ve{\varepsilon}
\def\vt{\vartheta}

\def\gs{\mathrel{\lower0.6ex\hbox{$\buildrel {\textstyle >}\over{\scriptstyle \sim}$}}}
\def\ls{\mathrel{\lower0.6ex\hbox{$\buildrel {\textstyle <}\over{\scriptstyle \sim}$}}}

\begin{document}

\title{Primary caustics and critical points behind a Kerr black hole}

\author{Mauro Sereno}
\email{sereno@physik.unizh.ch}

\author{Fabiana De Luca}

\affiliation{
Institut f\"{u}r Theoretische Physik, Universit\"{a}t Z\"{u}rich, Winterthurerstrasse 190, CH-8057 Z\"{u}rich, Switzerland
}

\date{February 24, 2008}

\begin{abstract}
The primary optical caustic surface behind a Kerr black hole is a four-cusped tube displaced from the line of sight. We derive the caustic surface in the nearly asymptotic region far from the black hole through a Taylor expansion of the lightlike geodesics up to and including fourth-order terms in $m/b$ and $a/b$, where $m$ is the black hole mass, $a$ the spin and $b$ the impact parameter. The corresponding critical locus in the observer's sky is elliptical and a point-like source inside the caustics will be imaged as an Einstein cross. With regard to lensing near critical points,  a Kerr lens is analogous to a circular lens perturbed by a dipole and a quadrupole potential. The caustic structure of the supermassive black hole in the Galactic center could be probed by lensing of low mass X-ray binaries in the Galactic inner regions or by hot spots in the accretion disk. 
\end{abstract}

\pacs{95.30.Sf, 04.70.Bw, 98.62.Sb}

\maketitle

\section{Introduction}

Black holes (BHs) are among the most fascinating predictions of the general theory of relativity. Recent progresses in mass measurements of compact objects have supplied compelling evidence for their existence. Nearly every galaxy is supposed to contain a very massive BH at its center in the mass range from $10^{6}$ to $10^{9}~M_\odot$ \citep{na+qu05}. Despite strong hints that massive BHs mostly rotate rapidly, their spins are still unmeasured so that a full characterization of their properties is not possible. Analyses of luminosities and spectra of accretion disks around spinning BH are very promising tool, but the interpretation can be sometimes unclear due to uncertainties in the physics of the inflowing gas. The classical test of gravitational deflection of light rays passing near compact bodies can provide an alternative probe. In fact, the theoretical bases of lensing  are  well understood and, on the observational side, the supermassive BH supposed to be hosted in the radio source Sgr~A* in the Galactic center, with a mass of $\sim 3.6\times 10^6 M_\odot$ and at a distance of $7.6~\mathrm{Kpc}$ from the Earth \citep{eis+al05}, offers a very appealing target for future space- and ground-based experiments. 

Lensing by rotating BHs has been studied from the very beginning of the Kerr spacetime \citep{bo+li67}. Angular momentum first appears in gravitational lensing through terms $\sim  m a /b^2$ in the deflection angle. Up to this order, light deflection is well understood. Very different approaches  can be undertaken \citep{pi+ro77,ep+sh80,iba83,ba+ho04,bra86,dym86,gli99,kop97,ko+sc99,ko+ma02,as+ka00,ser03b} to show the degeneracy between a Kerr BH and a suitably displaced Schwarzschild lens \citep{asa+al03,we+pe07}.  Such  analyses can be  easily extended to general spinning mass distributions \citep{ser02,se+ca02,ser03,ser05,ser07} and show many interesting features in the magnification pattern and in the image properties \citep{ser03,ser05,we+pe07} .

Investigations up to higher orders require the full consideration of the lightlike geodesics \citep{car68,cha83}.  The optical structure of the primary caustic surface \cite{ra+bl94}  as well as the appearance of both stars \citep{cu+ba73} and accretion disk \citep{vie93,be+do05} orbiting a Kerr BH have been detailed through numerical investigations. A clear analytical picture of the relativistic caustics in the strong deflection limit has also emerged \citep{boz+al05,bo+sc07}. The missing part, which we are going to provide here, is an analytical treatment of map singularities and caustics in the weak deflection limit. This is the prerequisite for the study of critical points and lensing map inversion near them.

The paper is organized as follows. In Section~\ref{sec:geod}, we recall basic properties of lightlike geodesics in Kerr spacetime. Sections~\ref{sec:caus} discusses the singularities of lens mapping and lensing near caustics. Section~\ref{sec:disc} is devoted to some final considerations. In this paper, we will use units $G=c=1$, with $c$ the light speed in the vacuum, so that the constant $m (\equiv G M /c^2)$ is the gravitational radius.

\section{Geodesic equations}
\label{sec:geod}

The Kerr black hole metric in the Boyer-Lindquist coordinates, $\{ t, r, \theta, \phi\}$, is given by
\ba
\label{ker1}
ds^2  &  = & \left( 1- \frac{2 m r}{\rho^2}\right)dt^2-\frac{\rho^2}{\Delta}dr^2 -\rho^2 d \theta^2 \\
& - &
\left( r^2 +a^2 +\frac{2 a^2 m r \sin^2 \theta }{\rho^2}\right)\sin^2 \theta d\phi^2   \nonumber \\
&  + & \frac{2 a m r \sin^2 \theta}{\rho^2}dt d\phi ,\nonumber
\ea
where
\begin{eqnarray}
\rho^2 & \equiv & r^2 +a^2  \cos^2 \theta , \nonumber \\
\Delta & \equiv & r^2 + 2 m r+a^2 . \nonumber
\end{eqnarray}

The null geodesics for a light ray can be expressed in terms of the first integrals of motion $J$ and $Q$, which are related to the impact parameter \cite{car68,cha83}. We consider photon trajectories from a source $S$, with Boyer-Lindquist coordinates $\{r_\mathrm{s}, \theta_\mathrm{s}, \phi_\mathrm{s}\}$, to an observer $O$ in $\{r_\mathrm{o}, \theta_\mathrm{o}, \phi_\mathrm{o}=0\}$. The photon trajectory can be expressed as
\begin{eqnarray}
\pm \int \frac{d r}{\sqrt{R}} & = & \pm \int \frac{ d \theta}{\sqrt{\Theta}} \label{ker4} , \\
-\phi_\mathrm{s} & = &\int \frac{a \left(2 mr-a J\right )}{\pm \Delta{\sqrt{R}} }dr +  \int \frac{ J d \theta}{\pm \sin^2 \theta \sqrt{\Theta} } \label{ker5} ,
\end{eqnarray}
where
\begin{eqnarray}
R(r) &  \equiv & r^4+\left(a^2-J^2-Q \right) r^2 + 2 \left[ (J-a)^2+Q\right] m r-a^2 Q , \nonumber\\
\Theta (\theta) & \equiv & \left(a^2-J^2 \csc ^2 \theta \right) \cos ^2\theta +Q .  \nonumber
\end{eqnarray}
The signs of $\sqrt{R}$ and $\sqrt{\Theta}$ are adhered to the signs of $d r$ and $d\theta$, respectively.  The geodesic equations in Eqs.~(\ref{ker4},\ref{ker5}) can be seen as lens equations, i.e. a mapping between the angular position of the source, that can be expressed through the corresponding Boyer-Lindquist coordinates, and the angular position of the images in the observer's sky, which are function of the constants of motion.

Within the standard framework of gravitational lensing, where the source of radiation and the observer are located in the nearly flat region of the spacetime, very far from , quantities of interest can be expanded in both $\epsilon_m \sim m/b$ and $\epsilon_a \sim a/b$.  In the nearly asymptotic region, $b/r_\mathrm{o} \sim b/r_\mathrm{s} \sim \epsilon_m$ \citep{ke+pe05}. In what follows, mixed terms like ${\cal{O}}(\epsilon_m^i) {\cal{O}}(\epsilon_a^j)$ are referred to as terms of order of ${\cal{O}}(\epsilon^{i+j})$ and we produce our results up to a given formal order in $\epsilon$.  

In previous works, the lightlike geodesic equations have been expanded as Taylor series up to and including second \citep{bra86} and third-order terms in $m/b$ and $a/b$ \citep{se+de06}, with the last analysis showing that the primary optical caustic is still point-like at that order.  Here, we take the further step and consider terms up to including contributions of order of $\sim \epsilon^4$. Calculations are performed strictly following the techniques developed in \citet{se+de06} . The equations of motion are approximated through an expansion of integrand functions and integrals. Radial and angular integrals are treated introducing the variables $\rho \equiv 1/r$ and $\mu \equiv \cos \theta$, respectively. Once performed the expansion of the geodesics equations in their integral form, our lens equations can be written as
\begin{eqnarray}
\phi_\mathrm{s} & = & \pm \pi  + \delta \phi_\mathrm{s} (J,Q) +{\cal{O}} (\epsilon^5), \\
\mu_\mathrm{s} & = & -\mu_\mathrm{o}  + \delta \mu_\mathrm{s} (J,Q)  +{\cal{O}}(\epsilon^5),
\end{eqnarray}
with $ \delta \phi_\mathrm{s}$ and $\delta \mu_\mathrm{s}$ denoting the deviation from the undeflected path. The constants of motion can be conveniently re-written in terms of $b_i$ parameters as \cite{se+de06},
$$
b_1   \equiv   - \frac{J}{\sqrt{1-\mu_\mathrm{o}^2}} , \ \
b_2   \equiv   - (-1)^k \sqrt{Q - J^2\frac{\mu_\mathrm{o}^2}{1-\mu_\mathrm{o}^2}} .  
$$
The parameter $k$ is an even (odd) integer for photons coming from below (above).

\begin{figure}
        \resizebox{\hsize}{!}{\includegraphics{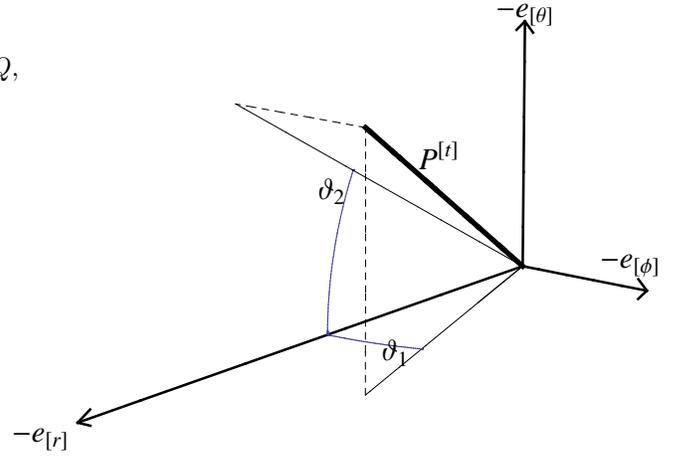}}
        \caption{The angles $\vt_1$ and $\vt_2$ in the locally flat observer's frame.} 
        \label{fig_vartheta_def}
\end{figure}

\begin{figure}
        \resizebox{\hsize}{!}{\includegraphics{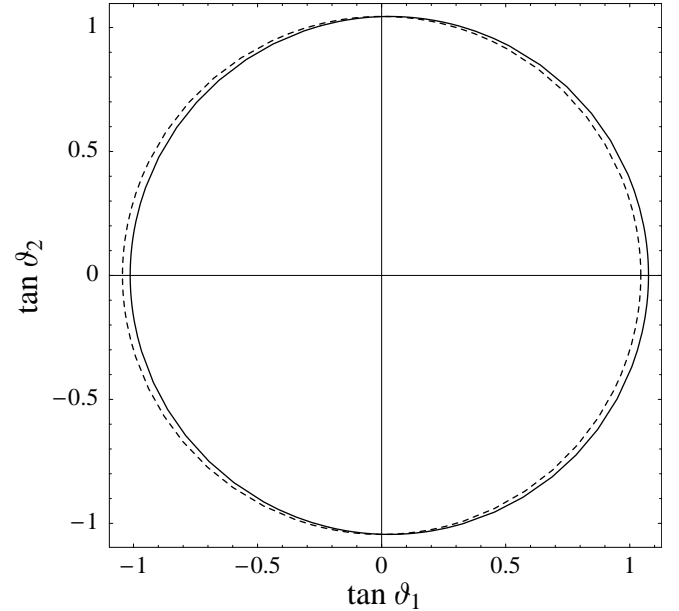}}
        \caption{Critical curve in the $\tan \vt_1$-$\tan \vt_2$ plane for a source at $r_\mathrm{s}=10~\mathrm{AU}$ behind a Sgr~A*-like BH. We consider an extreme Kerr, $a =m$ (full line),  and a Schwarzschild lens (dashed line).  The observer is equatorial, $\mu_\mathrm{o}=0$. Axis-lengths are in units of the tangent of the Einstein angle, $\theta_\mathrm{E} \equiv R_\mathrm{E}/r_\mathrm{o} \simeq  162~\mu\mathrm{arcsec}$.} 
        \label{fig_crit_10AU}
\end{figure}

\section{Caustics and critical points}
\label{sec:caus}

The critical points of the lens mapping form the locus of formally infinite magnification. They can be found where the Jacobian nulls out,
\beq
\left[ \frac{\partial \mu_\mathrm{s} \partial \phi_\mathrm{s}}{\partial b_1 \partial b_2} \right] = 0.  \label{magn1}
\eeq
Distances can be conveniently rescaled in terms of the Einstein ring $R_\mathrm{E} \equiv \sqrt{4  D r_\mathrm{o}  m  }$ and $D \equiv r_\mathrm{s}/(r_\mathrm{o}+r_\mathrm{s})$. A convenient new series expansion parameter is then $\varepsilon \equiv  R_\mathrm{E}/(4 D r_\mathrm{o})$ \citep{ke+pe05,se+de06}. Image positions near critical points can be written as a perturbation of the Einstein ring,
\ba
b_1 &  \simeq & R_\mathrm{E} \cos \varphi \left\{ 1 + {^{(1)}b_\mathrm{E}} \ve +{^{(2)}b_\mathrm{E}} \ve^2+{^{(3)}b_\mathrm{E}} \ve^3 \right\},\label{crit1}  \\
b_2 & \simeq  & R_\mathrm{E} \sin \varphi \left\{ 1 +{^{(1)}b_\mathrm{E}} \ve + {^{(2)}b_\mathrm{E}} \ve^2+{^{(3)}b_\mathrm{E}}  \ve^3 \right\} ,\label{crit2} 
\ea
with $0 \le \varphi \le 2\pi$. We are again collecting terms in two expansion parameters $\varepsilon$ and $\varepsilon_a \equiv a_m \varepsilon$, where $a_m (\equiv a/m)$ is the spin parameter in units of the mass, up to a given formal order. We recall that $\varepsilon$ and $\varepsilon_a$ are of the same physical order only for a maximal (or nearly maximal) Kerr black hole, when $|a| \sim m$.

Using Eqs.~(\ref{crit1},~ \ref{crit2}), the expanded Jacobian near critical points takes the form
\begin{widetext}
\ba
J   & \simeq  &  \frac{\varepsilon^3}{R_\mathrm{E}^2}  \left\{  30 \pi -64 {^{(1)}b_\mathrm{E}}  + 64  a_m \cos \varphi  \sqrt{1-\mu _o^2}  \right\}  
+ \frac{\varepsilon^4}{R_\mathrm{E}^2} \left\{  256( 1+D -D^2) +\frac{225 \pi ^2}{8} -20 (9 \pi -8 {^{(1)}b_\mathrm{E}} ) {^{(1)}b_\mathrm{E}}-64 {^{(2)}b_\mathrm{E}}  \right.  \nonumber \\
& + &  
\left. 
10  a_m \cos \varphi  \left(21 \pi -32 {^{(1)}b_\mathrm{E}}\right) \sqrt{1-\mu _o^2} +  32 a_m^2 \left(2 - 3 \mu _o^2 + 3(1- \mu _o^2) \cos 2 \varphi  \right)  
\right\}  
+
 \frac{\varepsilon^5}{R_\mathrm{E}^2}  \left\{  \frac{15 \pi}{8} \left[ 679 -64 D(1-D) \right]   \right.
\nonumber \\
&+&
\left.
585 \pi  {^{(1)}b_\mathrm{E}}^2
-320 {^{(1)}b_\mathrm{E}}^3 
- \left[1024\left(3 - D + D^2 \right)  + \frac{675 \pi ^2}{4}  - 320 {^{(2)}b_\mathrm{E}}  \right]{^{(1)}b_\mathrm{E}}
-180 \pi  {^{(2)}b_\mathrm{E}} -64 {^{(3)}b_\mathrm{E}}
\right. \nonumber \\
& + &
 a_m \sqrt{1-\mu _o^2}  \cos \varphi 
 \left[ 
 225 \pi^2+512 \left(7- D +D^2\right) -1380 \pi  {^{(1)}b_\mathrm{E}} + 960   {^{(1)}b_\mathrm{E}}^2   -320 {^{(2)}b_\mathrm{E}}
 \right] \\
& + &
 a_m^2 \csc \varphi \left[ 
\frac{15 \pi}{32}  \left( 121 -277 \mu _o^2 + 4(29+2 \mu _o^2) \cos 2  \varphi - 237 ( 1-\mu _o^2) \cos 4 \varphi    \right) \right. \nonumber 
\\
& -&
\left. \left.
16{^{(1)}b_\mathrm{E}}  \left(3 -11 \mu_o^2 + 6 \cos 2 \varphi - 9  (1 - \mu_o^2) \cos 4 \varphi   \right)
\right] - 
32 \cot \varphi \csc \varphi a_m^3 \sqrt{1-\mu _o^2} \left[ 1 + 3 \mu _o^2 -  3 \cos  2 \varphi  + 2(1 - \mu _o^2 ) \cos 4 \varphi   \right] \nonumber
\right\} .
\ea
\end{widetext}
The above series begins with terms of order of $\varepsilon^3$ since Eqs.~(\ref{crit1},~ \ref{crit2}) already take into account that at the lowest order critical points are located on the unperturbed Einstein circle. We can then find the higher order corrections to the position of the critical points by nulling out the Jacobian order by order. We get
\ba
^{(1)}b_\mathrm{E} & = &  \frac{15 \pi  }{32} + a_m \sqrt{1-\mu _\mathrm{o}^2}  \cos  \varphi  ,  \label{crit3} \\
^{(2)}b_\mathrm{E} & = &  4 \left(1 +D -D^2 \right)- \frac{675 \pi ^2}{2048}    +   \frac{15\pi}{32}    a_m \sqrt{1-\mu _\mathrm{o}^2}    \cos \varphi   \nonumber     \\
              & + & \frac{1}{2} \left[ \sin ^2 \varphi \left(\mu _\mathrm{o}^2-1\right)-\mu _\mathrm{o}^2\right] a_m^2 , \label{crit4}   \\
^{(3)}b_\mathrm{E} & = & \frac{15\pi}{512}   \left(-\mu _\mathrm{o}^2 - 3 \cos  2 \varphi   \left(1- \mu _\mathrm{o}^2 \right)-5\right) a_m^2  \nonumber \\
              & + & \left[  8 \left(1 + D - D^2\right)   -\frac{225\pi ^2}{256}   \right] \cos  \varphi  \sqrt{1-\mu _\mathrm{o}^2} a  \nonumber \\
& +& \frac{15 \pi }{4}  \left( \frac{225 \pi ^2}{2048}  -\frac{153}{128} +D - D^2 \right)  .\label{crit5} 
\ea
Putting back these solutions in the lens equations, we obtain the caustic surface at $r_\mathrm{s}$ in Boyer-Lindquist coordinates,
\begin{eqnarray}
\phi_\mathrm{s} & = & -\frac{b_1}{|b_1|} \pi  -4 a_m \varepsilon^2 -\frac{5\pi}{4} a_m \varepsilon^3  \label{caus1}  \\
 & + &  
\left\{ \left( \frac{225\pi^2}{128} -16 \right) a_m -\frac{15 \pi}{16} a_m^2 \sqrt{1-\mu_\mathrm{o}^2} \cos^3 \varphi  \right\} \varepsilon^4 \nonumber \\
\mu_\mathrm{s} & = & -\mu_\mathrm{o}  -\left\{ \frac{15 \pi}{16} a_m^2 (1-\mu_\mathrm{o}^2)^{3/2} \sin^3 \varphi \right\} \varepsilon^4  \label{caus2}   .
\end{eqnarray}
The terms proportional to $ \cos^3 \varphi $ and  $\sin^3 \varphi$ trace the astroid caustic surface. The terms in Eq.~(\ref{caus1}) proportional to $a$ produce the displacement from the line of sight connecting observer and lens.

We have now to translate our results in the observer's frame. The frame of reference of the observer can be oriented parallely to the local flat three-space of the locally nonrotating frame (LNRF)  and the position angles of the images in the observer's sky can be expressed in terms of the tetrad components of the four momentum $P$ of the photon \citep{bar+al72,vie93}. We take the basis vector $-e_{[\theta]}$ as polar axis; the azimuth is zero on the line of sight, i.e. the axis  $-e_{[r]}$,  and counted positive over $-e_{[\phi]}$, see Fig.~\ref{fig_vartheta_def}. The angles $\vt_1$ and $\vt_2$ are then defined such that $\tan \vt_1 = -P^{[\phi]}/ P^{[r]}$, i.e. $\vt_1$ is the opposite of the azimuthal angle, and $\tan \vt_2 = P^{[\theta]}/ P^{[r]}$. For an observer in the LNRF, the relations between the angles in the celestial sky and the constants of motion of the photon trajectories can be written as
\begin{widetext}
\ba
\tan \vartheta_1 & = &\frac{b_1}{r_\mathrm{o}} \frac{ \left(1+ \frac{a^2 \mu _o^2}{r_\mathrm{o}^2}\right) \sqrt{ 1 -\frac{2 m}{r_\mathrm{o}} +\frac{a^2}{r_\mathrm{o}^2}} }{
\sqrt{ 
1-\left( \frac{b}{r_\mathrm{o}}\right)^2 \left(1-\frac{2 m}{r_\mathrm{o}}\right) +
\frac{4 m a \sqrt{1-\mu _o^2} }{r_\mathrm{o}^2} \frac{b_1}{r_\mathrm{o}}
+
\left( \frac{a}{r_\mathrm{o}}\right)^2 \left( 1+\frac{2 m}{r_\mathrm{o}} -\frac{b_1^2 \mu _o^2}{r_\mathrm{o}^2} - \frac{b_2^2}{r_\mathrm{o}^2} \right) 
}
\sqrt{
1 + \frac{\mu _o^2 a^4}{r_\mathrm{o}^4} +\left(1 + \mu_o^2+ \frac{2 m (1-\mu_o^2)}{r_\mathrm{o}} \right)  \frac{a^2}{r_\mathrm{o}^2}}
} , \nonumber \\
\tan \vartheta_2 & = &\frac{b_2}{|b_2|} \frac{  \sqrt{\frac{b_2^2}{r_\mathrm{o}^2}+\frac{a^2 \mu _o^2}{r_\mathrm{o}^2}}   \sqrt{ 1 -\frac{2 m}{r_\mathrm{o}} +\frac{a^2}{r_\mathrm{o}^2}}     }{
\sqrt{ 
1-\left( \frac{b}{r_\mathrm{o}}\right)^2 \left(1-\frac{2 m}{r_\mathrm{o}}\right) +
\frac{4 m a \sqrt{1-\mu _o^2} }{r_\mathrm{o}^2}\frac{b_1}{r_\mathrm{o}} 
+
\left( \frac{a}{r_\mathrm{o}}\right)^2 \left( 1+\frac{2 m}{r_\mathrm{o}} -\frac{b_1^2 \mu _o^2}{r_\mathrm{o}^2} - \frac{b_2^2}{r_\mathrm{o}^2} \right) 
}} .
 \nonumber
\ea
\end{widetext}
In the plane spanned by $\vt_1$ and $\vt_2$ and orthogonal to the line of sight, the critical curve is an ellipse, see Fig.~\ref{fig_crit_10AU}, with equation
\beq
( \tan \vt_1 -\Delta_1)^2 + (1-e)^2\tan^2 \vt_2 = \vt_\mathrm{R}^2 ;
\eeq
the shift $\Delta_1$, the minor semi-axis $\vt_\mathrm{R} $ and the ellipticity $e$, defined such that the major semi-axis can be written as $\vt_\mathrm{R} / (1-e)$, are, respectively,
\ba
\Delta_1 & = &  \frac{a \sqrt{1-\mu _\mathrm{o}^2}}{r_\mathrm{o}}  
\left\{  
1 + \frac{15 \pi }{32} \varepsilon   \right. \nonumber   \\
 & + & \left. \left[   4 \left( 2 \pm D + 4 D^2 \right)   -\frac{225 \pi ^2}{256} \right] \varepsilon ^2    \right\} , 
\nonumber \\
 \vt_\mathrm{R} & = &  \theta_\mathrm{E} \left\{  1 + \frac{15 \pi}{32} \varepsilon  + \left[ 4 \left( 1 + D^2 \right) -\frac{675 \pi ^2}{2048} \right] \varepsilon ^2   + \frac{15 \pi}{8}  \varepsilon ^3  \right.  \nonumber  \\ 
& \times  &  \left.  
\left[       D + 4 D^2  - \frac{9  \left( 272 -25 \pi ^2\right)}{1024}  - \frac{a_m^2}{8}   
   \left(1 +\frac{3}{4} \mu _\mathrm{o}^2 \right)   \right] \right\} ,
\nonumber \\
e & = & \frac{105 \pi}{256}   a_m^2 \left(1 - \mu _\mathrm{o}^2\right)  \varepsilon ^3  , \nonumber 
\ea
where the upper and the lower signs hold for a LNRF and for a static observer, respectively. When comparing the above results with previous ones obtained in the framework of either non-rotating \citep{ke+pe05} or spinning \citep{se+de06} BHs, we have to remind that in this paper we properly defined the angles in the observer's frame instead of resorting to approximate relations between the angles and the constants of motion.

The unperturbed Einstein ring for a Sgr~A*-like lens and a source at $r_\mathrm{s} \sim 100\mathrm{AU}$ ($1\mathrm{pc}$) is $\sim 510~\mu\mathrm{arcsec}$ ($2.3 \times 10^{-2}\mathrm{arcsec}$). The center shift is dominated by the correction $\propto a$ and is nearly independent of $r_\mathrm{s}$, $\Delta_1 \ls 4.9 (4.8)~\mu\mathrm{arcsec}$ for $a \ls m$. The ellipticity is $\ls 1.1 \times 10^{-6}$ ($1.2 \times 10^{-11}$) which yields a difference in the axis lengths $ \ls 5.4 \times 10^{-4}$ ($2.8 \times 10^{-7}$)~$\mu\mathrm{arcsec}$. Finally the contribution $\propto a^2$ to the  radius is $\ls 3.1 \times 10^{-4}$ ($1.6 \times 10^{-7}$)~$\mu\mathrm{arcsec}$.

\begin{figure}
        \resizebox{\hsize}{!}{\includegraphics{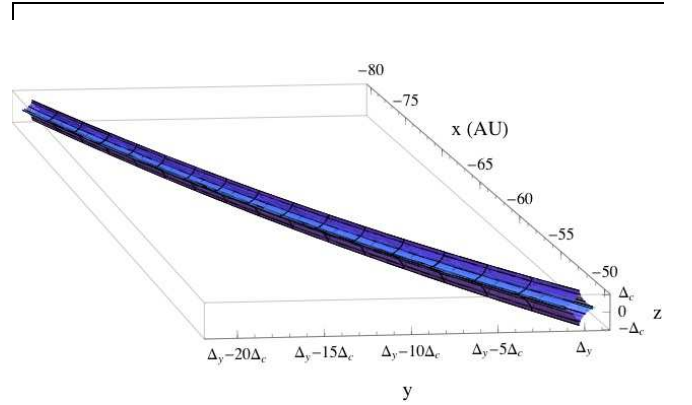}}
        \caption{Primary caustic surface between $r_\mathrm{s}=50~\mathrm{AU}$ and $r_\mathrm{s}=80~\mathrm{AU}$ behind a Sgr~A*-like lens. We consider extreme Kerr ($a=m$) and an equatorial observer ($\mu_\mathrm{o}=0$). In this distance range, the caustics drifts towards the line of sight and decreases in size. The $x$-axis length is in units of AU. The ticks on the $y$ and $z$-axes refer to the semi-width, $\Delta_\mathrm{c} \simeq 690~\mathrm{km}$, and to the displacement from the line of sight, $\Delta_\mathrm{y} \simeq 5.4\times 10^9~\mathrm{m}$, at $r_\mathrm{s}=50~\mathrm{AU}$.} 
        \label{fig_caus_50AU}
\end{figure}

\begin{figure}
        \resizebox{\hsize}{!}{\includegraphics{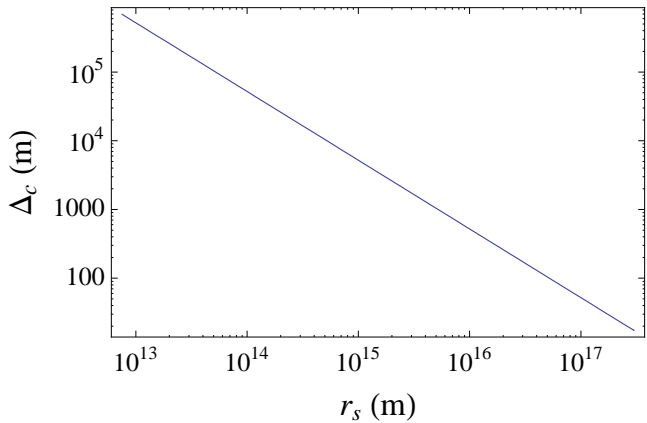}}
        \caption{The semi-width $\Delta_\mathrm{c}$ (in meters) of the caustic surface as a function of the radial source coordinate $r_\mathrm{s}$ behind a  Sgr~A*-like black lens. We consider $a=m$ and $\mu_\mathrm{o}=0$.} 
        \label{fig_Delta_c}
\end{figure}

\begin{figure}
        \resizebox{\hsize}{!}{\includegraphics{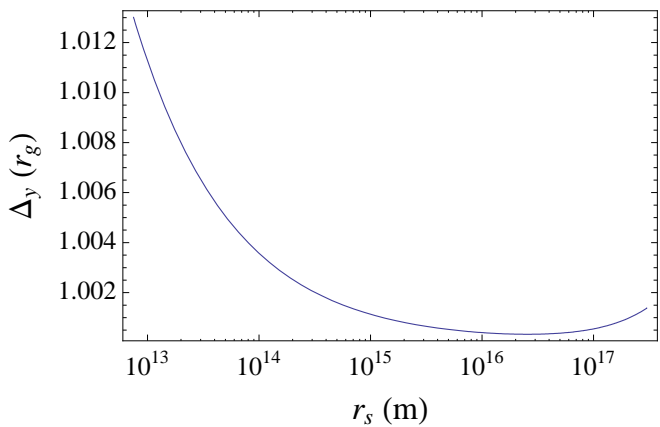}}
        \caption{The $y$-coordinate of the center of the caustic astroid as a function of the radial source coordinates $r_\mathrm{s}$ behind a  Sgr~A*-like black lens. The displacement $\Delta_\mathrm{y}$ is in units of the gravitational radius of the hole. We consider $a=m$ and $\mu_\mathrm{o}=0$.} 
        \label{fig_Delta_y}
\end{figure}

We can plot the caustic surface in a pseudo-Euclidean elliptical coordinate system related to the Boyer-Lindquist coordinates by $(x',y',z') =( \sqrt{r^2 + a^2} \sin \t \cos \phi,\sqrt{r^2 + a^2} \sin \t \sin \phi, r \cos \t)$. After such a transformation, for $m=0$ the Kerr metric reduces to the Minkowski one \citep{pl+kr06}. In a rotated system $(x,y,z)$ in which the $x$-axis lies along the line of sight, we get
\ba
x_\mathrm{s} & \simeq & -r_\mathrm{s} +  \frac{a^2  \left(1 - \mu _\mathrm{o}^2 \right)}{2D(1-D)r_\mathrm{o} } 
\left[  D (2-D)   + \frac{5 \pi}{8}\varepsilon   \right]   , \label{caus3}  \\
y_\mathrm{s}   & \simeq &  
\frac{a \sqrt{1-\mu _\mathrm{o}^2} }{1- D} \left[ 
1 + \frac{5 \pi }{16}  \varepsilon + \left(4 - \frac{225 \pi^2}{512}   \right)   \varepsilon ^2 \right]   \nonumber  \\
 &  +&   \Delta_\mathrm{c}  \cos ^3 \varphi    , \label{caus4}   \\
z_\mathrm{s} & \simeq &      -  \frac{a^2 \mu _\mathrm{o} \sqrt{1-\mu _\mathrm{o}^2}}{ (1-D) r_\mathrm{o}}  -\Delta_\mathrm{c} \sin ^3 \varphi   ,  \label{caus5} 
\ea
with
\beq
\Delta_\mathrm{c} =  \frac{15 \pi}{256}  \frac{ a^2 \left( 1- \mu _\mathrm{o}^2 \right) }{D (1-D) r_\mathrm{o}}.    \label{caus6} 
\eeq
The caustic surface  is a four-cusped astroid tube, see Fig.~\ref{fig_caus_50AU}.  The astroid is symmetric with total width equal to $2\Delta_\mathrm{c}$. The width decreases with increasing observer's distance $r_\mathrm{o}$ and decreases (increases) with increasing source distance if $r_\mathrm{o} > (<) r_\mathrm{s}$. For an asymptotic observer ($r_0 \rightarrow \infty$),  
\beq
\Delta_\mathrm{c} \simeq \frac{15 \pi}{256}  \frac{ a^2 \left( 1- \mu _\mathrm{o}^2 \right) }{r_\mathrm{s}},    \label{caus6bis} 
\eeq
in agreement with the numeric interpolation in eq. 6 in \citep{ra+bl94}, apart from a small difference in the overall numerical factor ($0.17$ instead of our $15 \pi/256 \sim 0.18$).  In Fig.~\ref{fig_Delta_c}, we plot the evolution of the caustic size behind a Sgr A*-like BH black hole, with the source coordinate $r_\mathrm{s}$ spanning the range from $50$~AU, slightly smaller than the pericentre of S2 (the observed orbiting star nearest to Sgr~A* \citep{eis+al05}), to $10$~pc, a distance slightly larger than the scale-length of the star cluster in the Galactic center. The width is really small. At $r_\mathrm{s} \sim 100\mathrm{AU}$ $(1\mathrm{pc})$, $\Delta_\mathrm{c} \ls 350~\mathrm{km}$ ($170~\mathrm{m}$) for $|a| \ls m$, a very tiny fraction of the gravitational  radius of the BH, $r_g \sim 5.3 \times 10^{6} \mathrm{km}$. 

The main contribution to the displacement from the line of sight is $\propto a$ in the $y$-direction, $\Delta_\mathrm{y} \sim a \sqrt{1-\mu_\mathrm{o}^2}/(1-D)$, and $\propto a^2$ along the $z$-axis. The caustic position drifts either towards or far away from the lens according to the relative observer and source position. For $m \ll r_\mathrm{s} \ll r_\mathrm{o}$, $ \Delta_\mathrm{y} \sim a \sqrt{1-\mu_\mathrm{o}^2}$ and the center moves towards the line of sight, whereas the trend is inverse at larger source distances. For $m \ll r_\mathrm{s} \sim r_\mathrm{o}$, $ \Delta_\mathrm{y} \sim 2 a \sqrt{1-\mu_\mathrm{o}^2}$. The above analysis of the caustic properties agrees with the numerical results in \cite{ra+bl94} and extend them in that it is not limited to very distant observers ($r_\mathrm{s} \ll r_\mathrm{o}$).

\subsection{Lensing inversion near caustics}
\label{subsec:inve}

As well known, the number of images of a source crossing the caustics changes by two. Let us consider a source displaced by $(\delta \phi_\mathrm{s}\ve^4, \delta \mu_\mathrm{s} \ve^4)$ with respect to the center of the caustic astroid. The solution for the image positions will be in the form of Eqs.~(\ref{crit1}), with $^{(3)}b_\mathrm{E}$ differing by $^{(3)}\delta b_\mathrm{E}$ from a point of the critical locus, Eq.~(\ref{crit5}). We get
\beq
\frac{15 \pi}{16}  (1-\mu _\mathrm{o}^2)^{3/2} a_m^2  \sin \varphi  + \delta \mu _\mathrm{s} -\delta \phi _\mathrm{s}  (1- \mu _\mathrm{o}^2) \tan \varphi =0  , \label{caus7} 
\eeq
and
\beq
^{(3)}\delta b_\mathrm{E} = \frac{15\pi}{128}  \left(1- \mu _\mathrm{o}^2 \right) a_m^2  \cos ^2 \varphi -\frac{\delta \phi _\mathrm{s} \sqrt{1-\mu _\mathrm{o}^2} }{8}   \sec  \varphi   .   \label{caus8} 
\eeq
The above expressions are typical of lensing near elliptical caustics and characterize also the strong deflection limit \cite{boz+al05,bo+sc07}. Equation~(\ref{caus7}) can be put in the form of a fourth degree polynomial equation in $\xi \equiv \sin \varphi$, so that a source has four images if inside the caustics and two if outside. The spin breaks the spherical symmetry. The Einstein ring does not form anymore and is replaced by an Einstein cross in the case of point-like source inside the caustics.  

The above results on critical points and on lensing near them show how there is a tight connection between a Kerr BH and a Schwarzschild lens slightly perturbed, as already envisaged in \citep{ra+bl94}. In fact, as far as the caustic structure is concerned, only the potential at the Einstein ring is important \citep{bl+ko88}. A first order dipole harmonic perturbation is enough to shift the point-like central caustics of a circular lens,  whereas the second harmonic yields the four-cusped asteroid. Such a quadrupole perturbation appears at order $(m a)^2/b^4$ in the deflection angle, breaking the degeneracy between Kerr and displaced Schwarzschild BH. Then, lensing distortion, magnification and formation of arcs by extended sources behind a Kerr lens follow the well known features of nearly circular gravitational lenses. The degeneration between a non null spin and other quadrupole perturbations, such as those caused either by deviations from spherically symmetry in the mass distribution of the lens or by external sources of shear, can be broken if the position of the deflector is determined with observative tools independent of lensing \cite{we+pe07}.

\section{Discussion}
\label{sec:disc}

The analytical derivation performed in this paper corroborates and extends knowledge about the primary caustics which, in a earlier work \citep{ra+bl94}, have been investigated only numerically and provides the first study, to our knowledge, of critical points and lens mapping near them. Our approach is the natural complement to qualitative methods which give some information on lensing properties without actually solving the equation for lightlike geodesics, such as the Morse theory \citep{ha+pe06}.

Together with the theoretical motivation, an equally compelling reason for investigating gravitational lensing in a Kerr spacetime comes from lensing observations towards Sgr~A* which are coming into the reach of observability and for which the  weak-deflection limit approximation at the lowest order is not applicable. The stars surrounding the Galactic center have been considered as suitable targets for detection of lensing effects. Sgr~A* is expected to be lensing nearly ten sources at any given time for observations down to $K \sim 21$ \citep{jar98,al+st99,cha+al01} . Considering the only few stars whose orbital parameters have already been accurately determined, detectable lensing events are expected to occur in a temporal span of $\sim 30$~years \citep{bo+ma05}. The typical  radius of these sources ($\sim 10^{8}-10^{9}~\mathrm{m}$) is much larger than the caustic width, so that the effects of the finite astroid size are washed out. Nevertheless, the astrometric shift of the caustic center, together with the precession orbital effects related to the spinning central body, must be accounted for when considering future experiments.

Other appealing sources for lensing by Sgr~A* are low mass X-ray binaries (LMXBs), whose emission mostly originates in a region a few tens of kilometers across, consisting of the inner accretion disk around the BH accreting from the companion. Tens of thousands of stellar-masses BHs and neutron stars are likely to have settled dynamically into the central parsec of the Galaxy  and perhaps few hundreds of them might have stellar companions \cite{mun+al05}. Several tens of LMXBs have been already detected in the very inner regions down to a minimum projected distance of only $0.1~\mathrm{pc}$ \citep{mun+al05b}, with a detected overabundance of transients X-ray binaries within $1~\mathrm{pc}$ \citep{mun+al05}.  These sources have been considered for detection of relativistic images \citep{boz+al05}  and could as well, since their small radius, probe the primary caustic surface.

Compact emission regions in the clumpy and unsteady accretion flow near Sgr~A* are other interesting lensing sources \citep{br+lo05,br+lo06}.  Relativistic images of orbiting bright-spots could be detected in the near future with observations achieving submilliarcsecond resolution at infrared and submillimetre wavelengths. However, we remark that in our approximation the source distance must be large and close inner orbits can not be considered. 

The finite size of the primary caustics behind Sgr~A* could then be probed by lensing of either LMXBs or compact hot spots in the accretion flow whereas the effects of angular momentum, such as the shift in the caustic position, affect even lensing of main sequence stars in the Galactic center cluster. 

\begin{acknowledgments}
M.S. is supported by the Swiss National Science Foundation and by the Tomalla Foundation. F.D.L. acknowledges the Forschungskredit of the University of Z\"{u}rich  for financial support.
\end{acknowledgments}


\end{document}